\documentclass{article}
\usepackage{graphicx}
\usepackage{amsmath}

\input{tcilatex}

\begin{document}

\title{Quantum walk on the line as an interference phenomenon}
\author{Peter L. Knight\thanks{%
E-mail: p.knight@imperial.ac.uk}, Eugenio Rold\'{a}n\thanks{%
Permanent address: Departament d'\`{O}ptica, Universitat de Val\`{e}ncia,
Dr. Moliner 50, 46100--Burjassot, Spain. E-mail: eugenio.roldan@uv.es}, and
J. E. Sipe\thanks{%
Permanent address: Department of Physics, University of Toronto, Toronto M5S
1A7, Canada. E-mail: sipe@physics.utoronto.ca} \\
Optics Section, Blackett Laboratory, Imperial College London, \\
London SW7 2AZ, United Kingdom}
\maketitle

\begin{abstract}
We show that the coined quantum walk on a line can be understood as an
interference phenomenon, can be classically implemented, and indeed already
has been. The walk is essentially two independent walks associated with the
different coin sides, coupled only at initiation. There is a simple analogy
between the evolution of walker positions and the propagation of light in a
dispersive optical fiber.
\end{abstract}

The quantum random walk (QW) was first proposed ten years ago by Aharonov 
\textit{et al.} \cite{Aharonov} as the quantum analog of the classical
random walk (RW). QWs are receiving much attention \cite{Farhi}-\cite{Kempe2}%
: as some problems are best solved in classical computation with algorithms
based on RWs, it is expected that these type of problems could be solved
even faster in a quantum computer. Preliminary investigations focused on the
nature of the QWs themselves. For example, Kempe \cite{Kempe} has shown that
the hitting time of the discrete QW from one corner of an $N$-bit hypercube
to the opposite corner is polynomial in the number of steps, $n$, whilst it
is exponential in $n$ the classical case. Subsequently Shenvi\textit{\ et al.%
} \cite{Shenvi} showed that a QW can perform the same tasks as Grover's
search algorithm, and Childs \textit{et al.} \cite{Childs} introduced an
algorithm for crossing a special graph exponentially faster that can be done
with a classical RW. Kempe \cite{Kempe2} has recently reviewed the field.

In the classical RW on the line, the ``walker'' (the particle or system
performing the RW) randomly takes one step to the right or to the left
depending on the result of tossing a coin. After $n$ steps, the probability
of finding the walker at a distance $m$ from the origin is given by the
binomial distribution, a Gaussian for large $n$ with a standard deviation $%
\sigma =\sqrt{n}$. In the QW, the role of the coin is played by a qubit (as, 
\textit{e.g.} a two--level atom or a spin--1/2 particle). As its classical
counterpart, the quantum walker moves to the right or to the left depending
on the internal state of the qubit. After each displacement, the state of
the qubit is set to a superposition state by means of a suitable unitary
transformation, typically a Hadamard, that plays the role of the toss of the
coin in the RW. Yet the QW is not a\textit{\ random} walk, as its time
evolution is completely deterministic. The probability distribution in the
quantum case is very different from the classical one: it resembles the Airy
function (Fig.1) and has a standard deviation that is linear with $n$. This
is the \textit{discrete} time QW that should be distinguished from the 
\textit{continuous} time QW \cite{Farhi,Kempe2}, which we will not consider
here.

Possible implementations of the QW have been proposed by a number of authors 
\cite{Travaglione,Dür,Sanders}. Here we show that a \textit{classical}
implementation of the QW is possible, in analogy with other processes
usually associated with quantum computing \cite
{Spreeuw,Cerf,Kwiat,Bhattacharya}. Indeed, we point out that a classical
implementation very similar to the one we are proposing has actually been
implemented by Bouwmeester \textit{et al.}\cite{Bouwmeester}, in the context
of the optical Galton board, without the authors explicitly noting this.
Other classical (interferometric) implementations of the QW have been
proposed recently \cite{Hillery,Jeong}, but in them the number of necessary
optical elements grows quickly with the number of steps in the QW, something
that does not occur in our scheme. Finally, by re-examining the difference
equations for the walker we show that the nature of propagation is simpler
than has been previously appreciated.

In our classical approach the role of the walker is played by the frequency
of a light field, and the role of the coin is played by its polarization
state. The light field 
\begin{equation}
\vec{E}=\sum_{m=-l}^{l}\vec{E}_{m}\exp \left[ i\left( \omega _{0}+m\bar{%
\omega}\right) t-ik_{m}z\right] +c.c.,
\end{equation}
($\omega _{0}$ is the carrier frequency and $\bar{\omega}$ is the frequency
difference between successive frequency components), can be represented by
the abstract state 
\begin{equation}
\left| \psi \right) =\sum_{m=-l}^{l}\left[ R_{m}\left| m,x\right)
+L_{m}\left| m,y\right) \right] ,
\end{equation}
where $R_{m}\equiv \hat{x}\cdot \vec{u}_{m}$ and $L_{m}\equiv \hat{y}\cdot 
\vec{u}_{m}$ ($\vec{u}_{m}=\vec{E}_{m}/\left| \vec{E}_{m}\right| $) and $%
\sum_{m=-l}^{l}\left[ \left| R_{m}\right| ^{2}+\left| L_{m}\right| ^{2}%
\right] =1$; the ``basis vectors'' $\left| m,c\right) $ label the frequency
and polarization, with $c=x,y$; we associated $x$($y$) with the coin head
(tail).

To implement the walk, we require a unitary operator that performs $\hat{V}%
\left| m,\QATOP{x}{y}\right) =\left| m\pm 1,\QATOP{x}{y}\right) $. The
operation $\hat{V}$ can be physically implemented, \textit{e.g.}, with an
electrooptic modulator (EOM) to which a linearly time dependent voltage is
applied in such a way that the $x$ ($y$) polarization component of the field
frequency component $\left( \omega _{0}+m\bar{\omega}\right) $ will see its
frequency increased (decreased) by an amount $\bar{\omega}$.

After each jump in the frequency of the field, a Hadamard transformation, $%
\hat{H}\left| m,\QATOP{x}{y}\right) =\frac{1}{\sqrt{2}}\left[ \left|
m,x\right) \pm \left| m,y\right) \right] $ has to be implemented. This can
be done optically by means of a half--wave plate (HWP) with its fast axis
forming an angle $\pi /8$ with respect to the $\hat{x}$ axis \cite
{Spreeuw,Cerf}. Finally, the QW is implemented by the repeated action on the
state of the operator $\hat{H}\hat{V}$, \textit{i.e.} after $n$ iterations $%
\left| \psi (n)\right) =\left[ \hat{H}\hat{V}\right] ^{n}\left| \psi
(0)\right) $, that can be written 
\begin{align}
\left| \psi (n)\right) & =\sum_{m=-N}^{+N}\left[ R_{m,n}\left| m,x\right)
+L_{m,n}\left| m,y\right) \right] \\
R_{m,n}& =\frac{1}{\sqrt{2}}\left( R_{m-1,n-1}+L_{m+1,n-1}\right) ,
\label{Rincrement} \\
L_{m,n}& =\frac{1}{\sqrt{2}}\left( R_{m-1,n-1}-L_{m+1,n-1}\right) ,
\label{Lincrement}
\end{align}
where $R_{m,0}=L_{m,0}=0\;$if$\;m\neq 0$ and $R_{m,-1}=L_{m,-1}=0\ \forall
m. $ These are the standard QW equations. Finally, the intensity of each
frequency component of the light field, which is the optical analog of the
probability of finding the walker at position $m$ at iteration (time) $n$,
is given by $P_{m,n}=\left| R_{m,n}\right| ^{2}+\left| L_{m,n}\right| ^{2}$,
which is represented in Fig.1.

In order to implement $n$ steps the best option is to introduce the
described elements in an optical cavity, Fig.2. The cavity imposes a
constraint that the optical frequencies must fit within its set of
eigenfrequencies. Thus, the time dependent electric field applied to the EOM
and the cavity length must be adjusted in such a way that the frequency
shift $\bar{\omega}=f\omega _{FSR}$ with $\omega _{FSR}$ the cavity free
spectral range and $f$ an integer number. Consider, \textit{e.g.}, that a
light pulse with a spectral width $\Delta \omega $ is initially injected in
the cavity. Then, in order to perform a step of the QW at each cavity
roundtrip, the step size $\bar{\omega}$ must be large enough to avoid
significant overlap between the spectrum of the displaced pulses thus the
frequency steps being well resolved. The total number of steps that can be
made in this QW then depends on both cavity losses and EOM bandwidth.

The experiment of Bouwmeester \textit{et al.} \cite{Bouwmeester} can be seen
as a realization of the QW very similar to the one proposed here. These
researchers proposed and studied, both theoretically and experimentally, an
optical implementation of the Galton Board (the quincunx). What they
actually implement is a grid of Landau-Zener crossings through which a light
beam propagates, and concentrate on the study of recurrences in the light
spectrum. A simplified version of their experimental device is that
represented in Fig.2, but with the QWP replaced by a second EOM with its
axis rotated $\pi /4$ with respect to the first EOM, which introduces a
dephasing between the two polarization components. Although this unitary
operation does not correspond to a Hadamard transformation, it can be shown
that it leads to an essentially identical QW \cite{Shenvi} (details to be
reported elsewhere). The main difference with our proposal is that the
frequency shift introduced by the EOM is \textit{smaller} than $\omega _{FSR}
$ and then each step in the QW takes several cavity roundtrips. In Fig.6 of
their paper \cite{Bouwmeester} the QW is clearly seen. Bouwmeester \textit{%
et al.} \cite{Bouwmeester} considered this case as a demonstration of the
coherence quality of their system, and did not note its significance to QWs;
their focus on the observation of recurrences in the spectrum led them to
study other aspects of their system.

Let us now re--examine the linear difference difference equations (\ref
{Rincrement},\ref{Lincrement}). They admit a formal solution that has been
studied from a number of points of view, usually with a focus on identifying
its asymptotic behavior for large $n$ \cite{Nayak,Carteret,Kempe2} as it
allows for the extraction of much information. Nevertheless the formal
solutions presented to date do not rely explicitly on a crucial feature of
Eqs. (\ref{Rincrement},\ref{Lincrement}), which we now explain, that greatly
simplifies a physical understanding of their solution. A little algebra
reveals that the solutions $R_{m,n}$ and $L_{m,n}$ of (\ref{Rincrement},\ref
{Lincrement}) also satisfy 
\begin{equation}
a_{m,n+1}=a_{m,n-1}+\frac{1}{\sqrt{2}}\left[ a_{m-1,n}-a_{m+1,n}\right]
,\;\;\;a=R,L.  \label{amp3}
\end{equation}
This is a remarkable equation, since it demonstrates a dynamical
independence of the evolution of the two coin states $R$ and $L$. Thus there
are two essentially independent walks, coupled only by the first step that
links the $a_{m,1}$ to the $a_{m,0}$. After that the two walks can be
studied independently of each other.

The most na\"{i}ve continuous limit of (\ref{amp3}) would involve a first
derivative with respect to time and a first derivative with respect to
space, and would suggest waves propagating only towards $+\infty$ for both $%
R $ and $L$, in apparent violation of the symmetry of the problem. But this
is too simplistic, given that for both $R$ and $L$ one can look for
solutions of the form,

\begin{equation}
a_{m,n}=A_{m,n}^{+}+\left( -1\right) ^{n}A_{m,n}^{-}  \label{ansatz}
\end{equation}
where $A_{m}^{\pm }$ satisfy 
\begin{equation}
A_{m,n+1}^{\pm }-A_{m,n-1}^{\pm }=\pm \frac{1}{\sqrt{2}}\left[
A_{m-1,n}^{\pm }-A_{m+1,n}^{\pm }\right] ,  \label{amp4}
\end{equation}
restoring the symmetry. Of course, there is not a unique specification of
the $A_{m}^{\pm }$ in terms of the fundamental $a_{m}$, since 
\begin{align}
a_{m,0}& =A_{m,0}^{+}+A_{m,0}^{-},  \label{atoA} \\
a_{m,1}& =A_{m,1}^{+}-A_{m,1}^{-}.  \notag
\end{align}
The specification of the $a_{m,0}$ and $a_{m,1}$, which completely specifies
the initial conditions required to solve (\ref{amp3}), does not suffice to
determine the initial conditions $A_{m,0}^{\pm }$ and $A_{m,1}^{\pm }$
required for the solution of (\ref{amp4}) uniquely. Nonetheless, it is
possible to rigorously develop the solution of the equations (\ref{amp3}) in
terms of the fields $A_{m,n}^{\pm }$; this we defer to a later publication.
The point we wish to stress here is that in the limit of $A_{m,n}^{\pm }$
that are slowly varying in $n$ and $m$ we can introduce continuous functions 
$A^{\pm }(x,t)$ and understand (\ref{amp4}) as the discretization of the
differential equation 
\begin{equation}
\sum_{k=0}^{\infty }\frac{\left( \Delta t\right) ^{2k+1}}{\left( 2k+1\right)
!}\frac{\partial ^{2k+1}}{\partial t^{2k+1}}A^{\pm }\left( x,t\right) =\mp 
\frac{1}{\sqrt{2}}\sum_{k=0}^{\infty }\frac{\left( \Delta x\right) ^{2k+1}}{%
\left( 2k+1\right) !}\frac{\partial ^{2k+1}}{\partial x^{2k+1}}A^{\pm
}\left( x,t\right) ,  \label{diff1}
\end{equation}
where $\Delta t$ and $\Delta x$ denote the temporal and spatial increments,
respectively. Keeping only the first two terms and approximating the third
derivative in time using the equation at the lowest order we obtain 
\begin{equation}
\frac{\partial }{\partial \tau }A^{\pm }\left( \xi ,\tau \right) =\mp \frac{1%
}{\sqrt{2}}\left[ \frac{\partial }{\partial \xi }+\frac{1}{12}\frac{\partial
^{3}}{\partial \xi ^{3}}\right] A^{\pm }\left( \xi ,\tau \right) 
\label{diff2}
\end{equation}
where $\tau =t/\Delta t$ and $\xi =x/\Delta x.$ In this slowly-varying
approximation the conditions (\ref{atoA}) can be approximated as 
\begin{equation}
A^{\pm }(m\left( \Delta x\right) ,0)\approx A_{m,0}^{\pm }=\frac{1}{2}\left(
a_{m,0}\pm a_{m,1}\right) ,  \label{iniA}
\end{equation}
where we have made use of Eq.(\ref{ansatz}) and assumed that $%
a_{m,1}=A_{m,1}^{+}-A_{m,1}^{-}\approx A_{m,0}^{+}-A_{m,0}^{-}$. Thus in
this limit (\ref{iniA}) provides the initial conditions for (\ref{diff2})
and those equations can be solved by Fourier analysis. There are then 
\textit{two }fields $A^{\pm }\left( \xi ,\tau \right) $ that can be
associated with each side of the coin. This feature persists when the
rigorous solution is constructed in this terminology, where there is a
(temporal) ``ferromagnetic'' field $A_{m,n}^{+}$ and an
``antiferromagnetic'' field $\left( -1\right) ^{n}A_{m,n}^{-}$ for each coin
side. 

Returning to Eq.(\ref{diff2}) we make use of Eqs.(\ref{iniA}) and take as
initial conditions $A^{^{\pm }}\left( \xi ,0\right) =a_{0,0}G\left( 0\right)
\pm a_{-1,1}G\left( -1\right) \pm a_{1,1}G\left( 1\right) $, with $G\left(
\xi _{0}\right) =\mathcal{N}\exp \left[ -\left( \xi -\xi _{0}\right)
^{2}/\left( 2\alpha \right) ^{2}\right] $ and $\mathcal{N}$ a normalization
factor; here we build in the fact that (\ref{diff2}) is only correct for the
long wavelength components by taking an initial condition that ``smears
out'' the lower wavelength components. The solution is easily found
analytically \cite{Miyagi,Abramowitz} for $A^{\pm }\left( \xi ,\tau \right) $%
, and we can write the final result (up to a normalization factor) for both $%
R$ and $L$ as $a\left( \xi ,\tau \right) =A^{+}\left( \xi ,\tau \right)
+\left( -1\right) ^{n}A^{-}\left( \xi ,\tau \right) ,$ with 
\begin{align}
A^{\pm }\left( \xi ,\tau \right) & =a_{0,0}Z\left( \pm \xi ,\tau \right)
\label{diff2solution} \\
& \pm a_{-1,1}Z\left( \pm \left( \xi +\xi _{0}\right) ,\tau \right) \pm
a_{1,1}Z\left( \pm \left( \xi -\xi _{0}\right) ,\tau \right) ,  \notag \\
Z\left( \xi ,\tau \right) & =\frac{2\pi }{\mathcal{B}^{1/3}}\exp \left( 
\frac{3\mathcal{ABC}+2\mathcal{C}^{3}}{3\mathcal{B}^{2}}\right) A_{i}\left( 
\frac{\mathcal{AB}+\mathcal{C}^{2}}{\mathcal{B}^{4/3}}\right) ,
\end{align}
where $\mathcal{A}=\xi -\tau /\sqrt{2}$, $\mathcal{B}=\tau /(4\sqrt{2})$, $%
\mathcal{C}=\alpha ^{2}$ and $A_{i}\left( x\right) $ is the Airy function 
\cite{Abramowitz}; the $R$ and $L$ solutions differ only through the
different values of $a_{m,0}$ and $a_{m,1}$ appearing in (\ref{diff2solution}%
). The appearance of Airy functions in the full solutions of (\ref
{Rincrement},\ref{Lincrement}) \cite{Carteret} can thus be understood as
associated with the form of equation (\ref{diff2}), which workers in fiber
optics will recognize as the classical equation for the propagation of light
in a fiber with no group velocity dispersion but a third-order dispersion
term. The linear dependence of the standard deviation on $n$ arises, of
course, simply because of this propagation. Solution (\ref{diff2solution})
is represented in Fig.1(b) for $\alpha =0.4$, and the similarity with the QW
in Fig.1(a) is clearly apparent.

In conclusion: We have shown that the QW along a line can be simulated in a
purely classical implementation, involving nothing more than wave
interference of electromagnetic fields. And, indeed, it has in fact already
been simulated in the laboratory in the work of Bouwmeester \textit{et al}. 
\cite{Bouwmeester}.

Further, this classical nature of the propagation is perhaps not surprising.
After all, the standard QW is a generalization of the quantum mechanical
problem of a spinless particle with hopping amplitudes between sites,
familiar from solid state physics if the time variable is continuous. That
latter problem, which gives a simple Schr\"{o}dinger equation in its
continuum limit, is clearly classical in the nature of its propagation, as
attested to by the appearance of the Schr\"{o}dinger equation in classical
beam propagation problems. The generalization involved in concocting the
standard QW problem is the inclusion of a spin variable. What we have shown
here is that this generalization does not affect the dynamics in an
essential way. Except for an initial coupling in the first two time steps,
the evolutions of the amplitudes associated with the two sides of the coin
proceed independently.

This work has been supported in part by the UK Engineering and Physical
Sciences Research Council and the European Union. ER acknowledges financial
support from the Ministerio de Educaci\'{o}n, Cultura y Deportes of the
Spanish Goverment (Grant PR20002-0244). JES acknowledges financial support
from the Natural Sciences and Engineering Research Council of Canada. We
gratefully acknowledge fruitful discussions with V. Kendon.

\bigskip

{\LARGE Figure Captions}

\textbf{Figure 1:} (a) Probability distribution for $n=200$ for both the
classical (dashed) and quantum (continuous) random walks. The initial
conditions chosen for calculating the QW were $R_{0,0}=1/\sqrt{2}$ and $%
L_{0,0}=i/\sqrt{2}$, see Eqs.(\ref{Rincrement},\ref{Lincrement}). Notice
that the quantum $P_{m}$ is null for odd $m$ at odd $n$. We have represented
only nonzero values. (b) Continuous limit of the QW as given by Eq.(\ref
{diff2solution}) for $\alpha=0.4$ and $t=200$ with the same initial
conditions as in (a).

\textbf{Figure 2:} Scheme for the optical implementation of the QW in a
Fabry-Perot cavity. The electrooptic modulator (EOM) shifts the field
frequency up or down in $\bar{\omega}/2$ depending on its polarization and a
quarter-wave plate (QWP) with its axis forming an angle $\pi /8$ with
respect to the $x$-axis, performs the Hadamard transformation (notice that
the light passes twice through each intracavity element every roundtrip).


\begin{thebibliography}{99}
\bibitem{Aharonov}  Y. Aharonov, L. Davidovich, and N. Zagury, Phys. Rev. A 
\textbf{48}, 1687 (1993)

\bibitem{Farhi}  E. Farhi and S. Gutmann, Phys. Rev. A \textbf{58}, 915
(1998)

\bibitem{Nayak}  A. Nayak, and A. Vishwanath, e-print quant-ph/0010117

\bibitem{Kempe}  J. Kempe, e-print quant-ph/0205083

\bibitem{Shenvi}  N. Shenvi, J. Kempe, and K.B. Whaley, Phys. Rev. A \textbf{%
67}, 052307 (2003)

\bibitem{Childs}  A.M. Childs, R. Cleve, E. Deotto, E. Farhi, S. Gutmann,
and D. A. Spielman, e-print quant-ph/0209131

\bibitem{Kendon}  V. Kendon and B. Tregenna, Phys. Rev. A \textbf{67},
042315 (2003)

\bibitem{Hillery}  M. Hillery, J. Bergou, and E. Feldman, e-print
quant-ph/0302161

\bibitem{Carteret}  H.A. Carteret, M.E.H. Ismail, and B. Richmond, e-print
quant-ph/0303105

\bibitem{Travaglione}  B.C. Travaglione and G.J. Milburn, Phys. Rev. A 
\textbf{65}, 032310 (2002)

\bibitem{Dür}  W. D\"{u}r, R. Raussendorf, V.M. Kendon, and H.-J. Briegel,
Phys. Rev. A \textbf{66}, 052319 (2002)

\bibitem{Sanders}  B.C. Sanders, S.D. Bartlett, B. Tregenna, and P.L.
Knight, Phys. Rev. A \textbf{67}, 042305 (2003)

\bibitem{Jeong}  H. Jeong, M. Paternostro, and M.S. Kim, e-print
quant-ph/0305008

\bibitem{Kempe2}  J. Kempe, Contemp. Phys. (to be published), e-print
quant-ph/0303081

\bibitem{Spreeuw}  R.J.C. Spreeuw, Phys. Rev. A \textbf{63}, 062302 (2001)

\bibitem{Cerf}  N.J. Cerf, C. Adami, and P.G. Kwiat, Phys. Rev. A \textbf{57}%
, R1477 (1998)

\bibitem{Kwiat}  P.G. Kwiat, J.R. Mitchell, P.D.D. Schwindt, and A.G. White,
J. Mod. Opt. \textbf{47}, 157 (2000)

\bibitem{Bhattacharya}  N. Bhattacharya, H.B. van Linden van den Heuvell,
and R.J.C. Spreeuw, Phys. Rev. Lett. \textbf{88}, 137901 (2002)

\bibitem{Bouwmeester}  D. Bouwmeester, I. Marzoli, G.P. Karman, W. Schleich,
and J.P. Woerdman, Phys. Rev. A \textbf{61}, 013410 (2000)

\bibitem{Miyagi}  M. Miyagi and S. Nishida, Appl. Opt. \textbf{18}, 678
(1979)

\bibitem{Abramowitz}  M. Abramowitz and I. A. Stegun, \textit{Handbook of
Mathematical Functions} (Dover, New York, 1970), pp.446,475
\end{thebibliography}
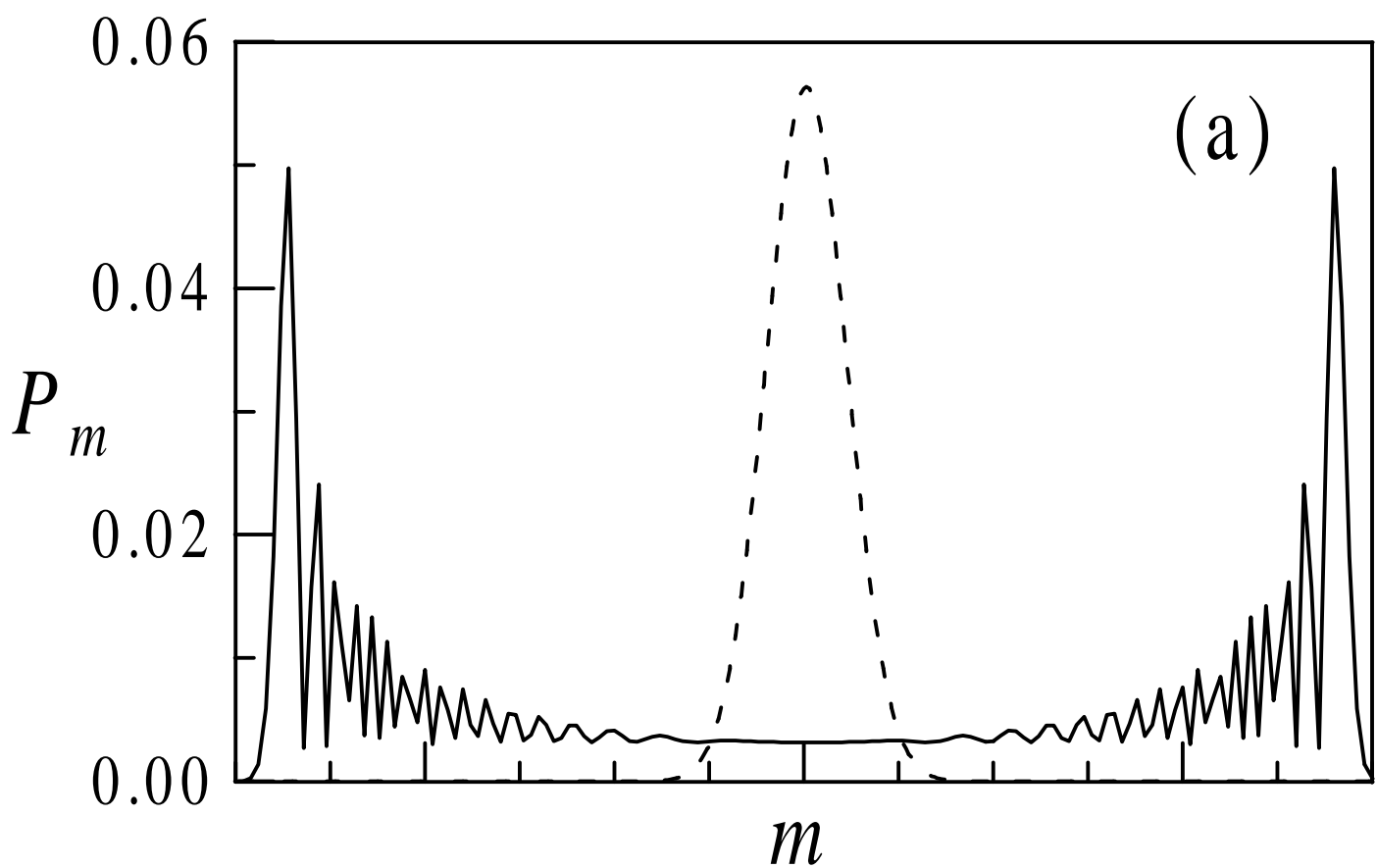
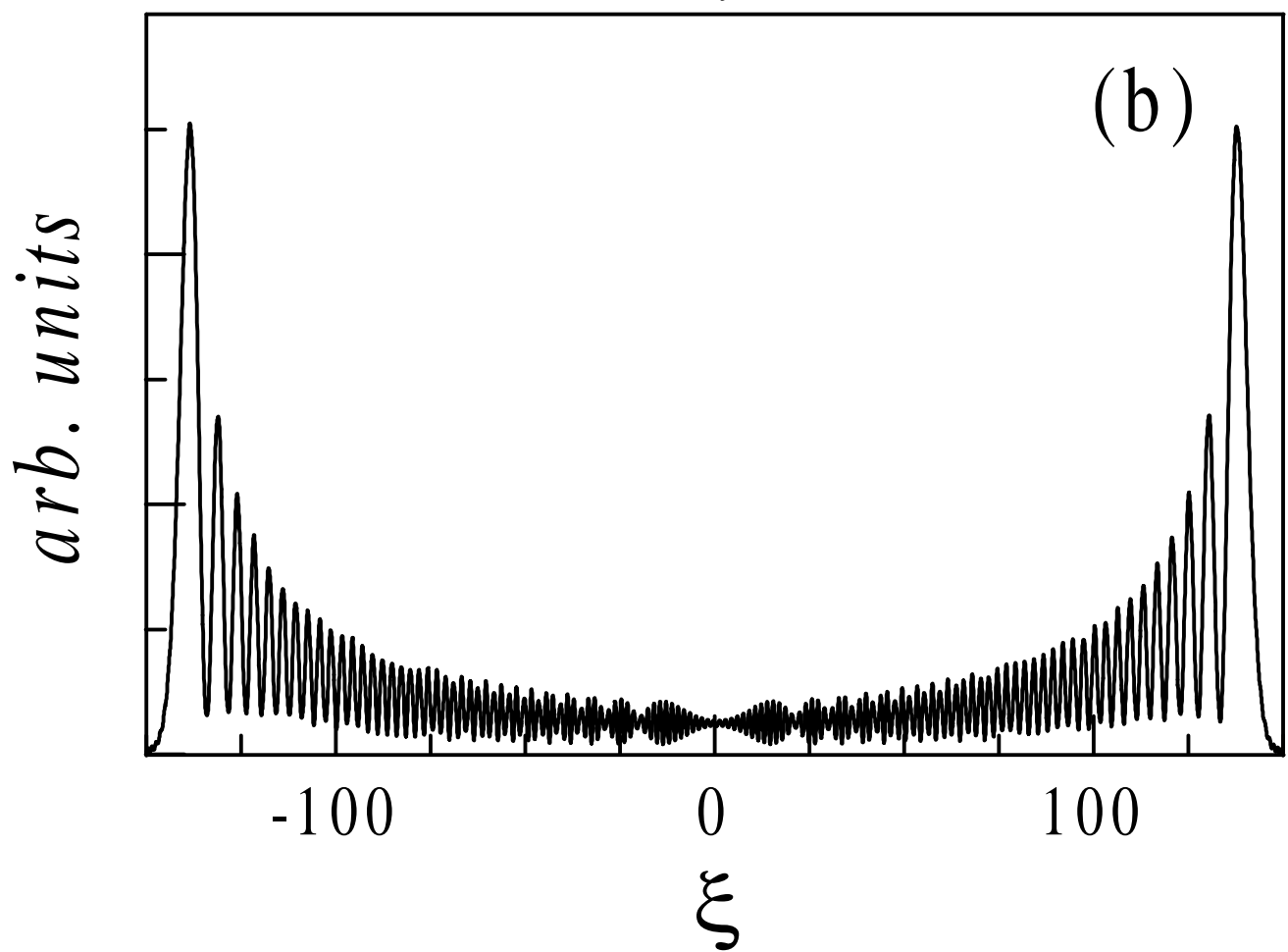
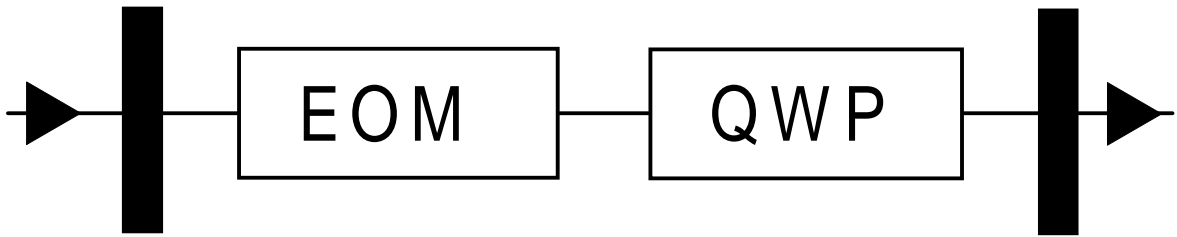
\end{document}